\def\year{2020}\relax
\documentclass[letterpaper]{article} 
\usepackage{aaai20_arxiv}  
\usepackage{times}  
\usepackage{helvet} 
\usepackage{courier}  
\usepackage[hyphens]{url}  
\usepackage{graphicx} 
\usepackage{graphicx}  
\usepackage{url}            
\usepackage{booktabs}       
\usepackage{amsfonts}       
\usepackage{nicefrac}       
\usepackage{lipsum}
\usepackage{subfigure}
\usepackage{amsmath}
\DeclareMathOperator*{\argmax}{arg\,max}

\title{Partner Selection for the Emergence of Cooperation in Multi-Agent Systems\\ Using Reinforcement Learning}
\author{
    Nicolas Anastassacos,\textsuperscript{\rm 1}\textsuperscript{\rm 2}
    Stephen Hailes,\textsuperscript{\rm 2}
    Mirco Musolesi\textsuperscript{\rm 1}\textsuperscript{\rm 2}\textsuperscript{\rm 3} \\ 
    \textsuperscript{\rm 1}The Alan Turing Institute
    \textsuperscript{\rm 2}University College London\\
    \textsuperscript{\rm 3}University of Bologna\\
}

\begin{document}

\maketitle

\begin{abstract}

Social dilemmas have been widely studied to explain how humans are able to cooperate in society. Considerable effort has been invested in designing artificial agents for social dilemmas that incorporate explicit agent motivations that are chosen to favor coordinated or cooperative responses. The prevalence of this general approach points towards the importance of achieving an understanding of both an agent's internal design and external environment dynamics that facilitate cooperative behavior. In this paper, we investigate how partner selection can promote cooperative behavior between agents who are trained to maximize a purely selfish objective function. Our experiments reveal that agents trained with this dynamic learn a strategy that retaliates against defectors while promoting cooperation with other agents resulting in a prosocial society. 

\end{abstract}

\section{Introduction}

Human history provides many examples of people behaving more cooperatively than is typically assumed in behavior models featuring assumptions of rationality and self-interest. Examples include behaving environmentally responsibly (littering, recycling, etc.) \cite{cialdini2003crafting,thogersen2006norms} or citizens honestly paying their taxes when there are financial incentives to do otherwise \cite{posner2000law}. These types of situations are known as social dilemmas and are characterized by a trade-off between an individual's short-term rewards and the collective long-term interests of the group, community or society as a whole \cite{dawes1980social}. Understanding the mechanisms for the emergence of cooperation is still an open problem in several disciplines ranging from evolutionary biology and economics to autonomous systems. 

Recently, Reinforcement Learning (RL) has been applied to the study of general-sum multi-agent games like social dilemmas \cite{hughes2018inequity,jaques2019social,leibo2017multi,perolat2017multi}, however, it is known that it is difficult to obtain optimal results due to the non-stationarity caused by agents learning simultaneously \cite{busoniu2008comprehensive}. Solutions to these problems can involve introducing techniques like modeling opponents or using joint-action distributions over the population \cite{lowe2017multi} though these techniques tend to be affected by scalability issues. 
The traditional RL objective describes a selfish agent that looks to maximize its own reward. In decentralized settings, it may not be possible to impose any cooperative reward function on all the agents. In these cases, it is important for the environment to be designed in a way that allows even selfish agents to learn behaviors which do not severely hinder other agents and, preferably, are beneficial for the society as a whole. We argue that an answer to training RL agents in decentralized multi-agent scenarios potentially lies in understanding the societal dynamics that produces \textit{norm-inducing
behaviors} and their effects on how and what agents learn.

We investigate a key characteristic of social interaction: partner selection. The capability for an individual to freely choose who they want to interact with has been thought to have a prominent role in determining the structure of a population and the competitive and collaborative relationships that form between members of society \cite{santos2008social}. In turn, these relationships can cause a change in the strategies that agents learn that may go on to impact partner selection in the future. The development is dynamic and cyclical and has been hypothesized to be a driving factor in the emergence of cooperative societies and, potentially, a catalyst for altruistic behavior \cite{barclay2006partner,cuesta2015reputation}.

Reputation and signaling are directly tied to the notion of partner selection. Individuals prefer cooperative partners and are inclined to avoid partnerships with known selfish individuals \cite{albert2007we,milinski2002donors}. This tendency towards preferring to interact with reputable individuals suggests that the freedom to choose may lead agents to develop strategies that maximize the reward, while also improving reputation or signaling in order to attract the attention of others. Signaling a desire to coordinate or cooperate, however, may come at the cost of becoming a target for exploitation by others. The interplay of these dynamics has made partner selection a key component of group formation and group inequality \cite{bolton2000erc,mohtashemi2003evolution}. Fu et al. developed a reputation-based partner switching model that predicts a cooperative response using agents \cite{fu2008reputation} and the repeated Prisoner's Dilemma; using result from evolutionary game theory they derive rules to facilitate partner switching in order to penalize agents with a low reputation and a network framework that determines the amount of information that is available to each agent. Motivated by these findings, we propose a model of partner selection that is characterized by two components: (1) $N$-agents are fairly presented with only the previous interactions of their opponents and use this information to select a partner, and then (2) each pair of agents engages in a Dilemma game and \textit{learns} from experience. Using this framework and RL agents trained with $Q$-learning \cite{watkins1992q}, we are able to simulate the emergence of cooperation in a decentralized society using partner selection and agents that learn according to a selfish objective function. When we contrast our results with the outcomes of scenarios in which agents are randomly paired, the impact of partner selection can be clearly seen on both the overall societal outcome and on the evolution of the strategies that are played throughout the simulation.

We show that when selfish agents have the option to choose a partner they are able to learn to cooperate with ``good'' individuals and punish ``bad'' ones, a strategy of equivalent retaliation known as Tit-for-Tat (TFT) \cite{axelrod1981evolution}. In our results, we can see how TFT agents act to sustain cooperation in a society by playing a strategy that encourages other agents to play the \textit{same} strategy akin to a social norm. Q-agents are capable of learning this strategy in the presence of partner selection and maintain cooperation in the society, whereas agents trained with random matching learn to always defect. The contributions of this work are two-fold. Firstly, we demonstrate the importance of building environments with societal dynamics like partner selection, such that agents can learn norm-inducing behaviors in decentralized multi-agent settings. Secondly, we provide novel insight into how cooperation can emerge between selfish agents that can be contrasted with outcomes seen in evolutionary biology and behavioral economics as we use familiar framework in the repeated Prisoner's Dilemma and a bottom-up approach to partner selection that results in a well-known strategy, TFT. For this reason, we believe that the results of this work have implications not only for the design of cooperative artificial agents but also for our understanding behavior in animal and human societies.

\section{Social Dilemmas}

Social dilemmas provide a powerful platform to study the emergence of behavior and the development of cooperative solutions using agents \cite{izquierdo2008reinforcement,macy2002learning,Sen2007}. 
 
They have featured heavily in behavioral economics, psychology and evolutionary biology 
\cite{axelrod1981evolution,mohtashemi2003evolution,fehr2004social,fu2008reputation,santos2008social}. Since then, other research has looked to explain cooperative outcomes in more complex environments termed sequential social dilemmas and investigate innate human-inspired characteristics such as a preference for fairness, pro-sociality, or social influence while leveraging the capabilities of RL agents to learn \cite{hughes2018inequity,jaques2019social,leibo2017multi,perolat2017multi,peysakhovich2018prosocial}. 

\begin{table}[t]
\centering
\begin{tabular}{c | c | c}
 SD & C & D \\
 \hline
 C & \textit{R, R} & \textit{S, T} \\
 \hline
 D & \textit{T, S} & \textit{P, P} \\
\hline
\end{tabular}
\quad \quad \quad
\begin{tabular}{c | c | c}

 rPD & C & D \\
\hline
 C & 3, 3 & 0, 4 \\
 \hline
 D & 4, 0 & 1, 1 \\
\hline
\end{tabular}
\quad \quad \quad
\caption{Payoff Matrix for Social Dilemmas and repeated Prisoner's Dilemma. The motivation to defect comes from fear of an opponent defecting or acting greedily to gain the maximum reward when one anticipates the opponent might cooperate. The game is modeled so that $T>R>P>S$ and $2R>T+S$. From a game theoretic perspective, the optimal strategy in a finite game is to defect. This is undesirable as the agents could both achieve greater reward if they agreed to cooperate.}
\label{table:social dilemma}
\end{table}

We consider a classic repeated Prisoner's Dilemma (PD) approach, where each iteration of the game can be characterized by a payoff matrix. Agents play multiple rounds of the Prisoner's Dilemma and there is continuity between games as an agent's previous actions are used to inform the partner selection of the other agents. The resulting dilemma is straightforward: agents must decide whether they should (1) maximize their immediate reward by defecting while broadcasting information to other agents that may deter them from future interactions or (2) cooperate, forgoing the immediate reward and risk being exploited, but potentially attract the attention of other agents to increase future returns. For clarity, we will name the outcomes in Table \ref{table:social dilemma} as follows: (C, C) is mutual cooperation, (D, C) is exploitation, (C, D) is deception, and (D, D) is mutual defection.

\subsection{Q-Learning}

RL algorithms learn a policy from experience balancing exploration of the environment and exploitation.

We train our agents with $Q$-learning \cite{watkins1992q}. We train agents independently using this algorithm in a multi-agent setting. In $Q$-learning, the policy of agent $i$ is represented through a state-action value function $Q^i(s, a)$. The $i$-th agent stores a function $Q_i: \mathcal{S}_i \times \mathcal{A}_i \rightarrow \mathbb{R}$ often parameterized by a neural network when dealing with a high dimensional state space. 

The policy of agent $i$ is an $\epsilon$-greedy policy and is defined by 
\begin{equation}
    \pi_i(s) = 
    \begin{cases}
        \argmax_{a \in \mathcal{A}_i} Q_i(s, a) & \text{with probability } 1-\epsilon \\
        \mathcal{U}(\mathcal{A}_i) & \text{with probability } \epsilon\\
    \end{cases}
\end{equation}

where $\mathcal{U}(\mathcal{A}_i)$ denotes a sample from the uniform distribution over the action space. Each agent stores a set of trajectories $\{(s, a, r_i, s')_t : t = 1,...,T\}$ by interacting with the environment and then updates its policy according to 

\begin{equation}
    Q_i(s,a) \leftarrow Q_i(s, a) + \alpha [r_i + \gamma \max_{a' \in \mathcal{A}_i} Q_i(s', a') - Q_i(s, a)]
\end{equation}

where $s$ is the current state, $a$ is the current action, $r_i$ is the reward obtained by agent $i$ and $s'$ is the next state. As agents are treated independently, the learning makes the environment appear non-stationary from any agent's perspective.
Other works have attempted to address the non-stationarity problem in multi-agent RL using a combination of joint-action learning and importance sampling methods \cite{foerster2017stabilising}. However, for large populations of agents this significantly increases the amount of required computation and becomes infeasible for large states and action spaces. In order to ensure that information stored in each agent's memory buffer is relevant to the current transition dynamics of the environment, the buffers are refreshed after every episode and the agent only trains only on the most recent experiences.

\subsection{Reinforcement Learning In Multi-Agent Social Dilemmas}

RL is a useful tool for understanding social dilemmas. In contrast with game theoretic models, deep RL does not assume knowledge of the environment dynamics and it is able to deal with continuous state and action spaces. It can therefore be used with more complex environmental dynamics that are more representative of the real-world such as investigating resource appropriation in the presence of scarce resources \cite{perolat2017multi} or cooperative hunting to maximize efficiency \cite{leibo2017multi}. Furthermore, it has also been used to analyze the outcome of agents trained with to be inequity averse with social dilemmas \cite{hughes2018inequity}, and of societies that feature agents with prosocial behavior \cite{peysakhovich2018prosocial} building on works in behavioral economics. 

The main difficulties in applying RL to multi-agent social dilemmas are credit-assignment, non-stationarity and incentive misalignment. The actual return of an agent's action is not contained just in the reward but is reflected in what the other agents learn. If an agent defects, the consequences of an action must be present in the trajectory in order for an agent to assign an appropriate value. It is difficult to capture this information in a trajectory without knowing when and how other agents are learning. To help account for this, the notion of intrinsic rewards or agent preferences is often utilized during training \cite{eccles2019learning}. This drives agents to learn coordinated behavior that emphasizes cooperation. In \cite{hughes2018inequity} agents are required to balance the external rewards they receive with their internal preference for more equal outcomes. In \cite{jaques2019social}, the reward that agents receive is also split into rewards that are received from the game environment but also rewards agents for taking actions that are measured to be ``highly influential'' based on the change it causes in other agents' behaviors. Other work has involved trying to predict changes in opponent's future behavior in response to current events \cite{foerster2018learning}. In contrast, in this work agents do not receive any rewards during training other than those received directly from playing the Dilemma.

\section{Modeling Methodology}

$N$ agents play rounds of a Dilemma game (Table \ref{table:social dilemma}). In our setup, agents are able to select a partner at the beginning of every round to play the Dilemma and the goal of the game is to achieve the highest possible individual reward after $T$ rounds. This also means that each agent takes at least $T$ actions. The best collective outcome for society is for all the agents to cooperate unconditionally as this would guarantee maximum reward for every interaction. However, such an outcome is unlikely because the immediate rewards associated with defecting in a highly cooperative society are also significantly higher. By introducing partner selection, agents have to balance the rewards of greedy defecting behavior with the future cost of being excluded. 

An episode of the game is split into two phases: in the first phase, each agent has a chance to select another agent to play the Dilemma. During this phase, actions $a_s \in \mathbb{R}^{n-1}$ involve selecting one of the other $n - 1$ agents in the environment. Agents cannot select themselves and, if selected, they cannot refuse to play. There are no other restrictions on which agents can be selected nor how many times any agent can be selected. In the second phase, the selecting agent, $i$, and chosen agent, $j$, play one round of the Dilemma, taking actions $(a^i_d, a^j_d)$ where $a_d \in [C, D]$ and receive reward $r_d$. Actions and rewards are according to the payoff in Table \ref{table:social dilemma}. Agents receive no reward from selecting a partner and only receive rewards when playing the Dilemma. Their rewards from playing must therefore inform their selection.

During selection, every agents' most recent $h$ actions are visible to all other agents. Each action is represented as a one-hot encoding and concatenated together such that for the playing phase agents see states $s_d \in \mathbb{R}^{2 \times h}$ and in the selection phase $s_s \in \mathbb{R}^{2\times h \times(n-1)}$. We choose to set $h=1$ for interpretability. This limits the number of strategies that can be learned to just four strategies that are recognized in other literature (see Results).

\begin{figure*}[!h]

  \subfigure[Matching determined by partner selection]
  {\includegraphics[width=0.95\columnwidth]{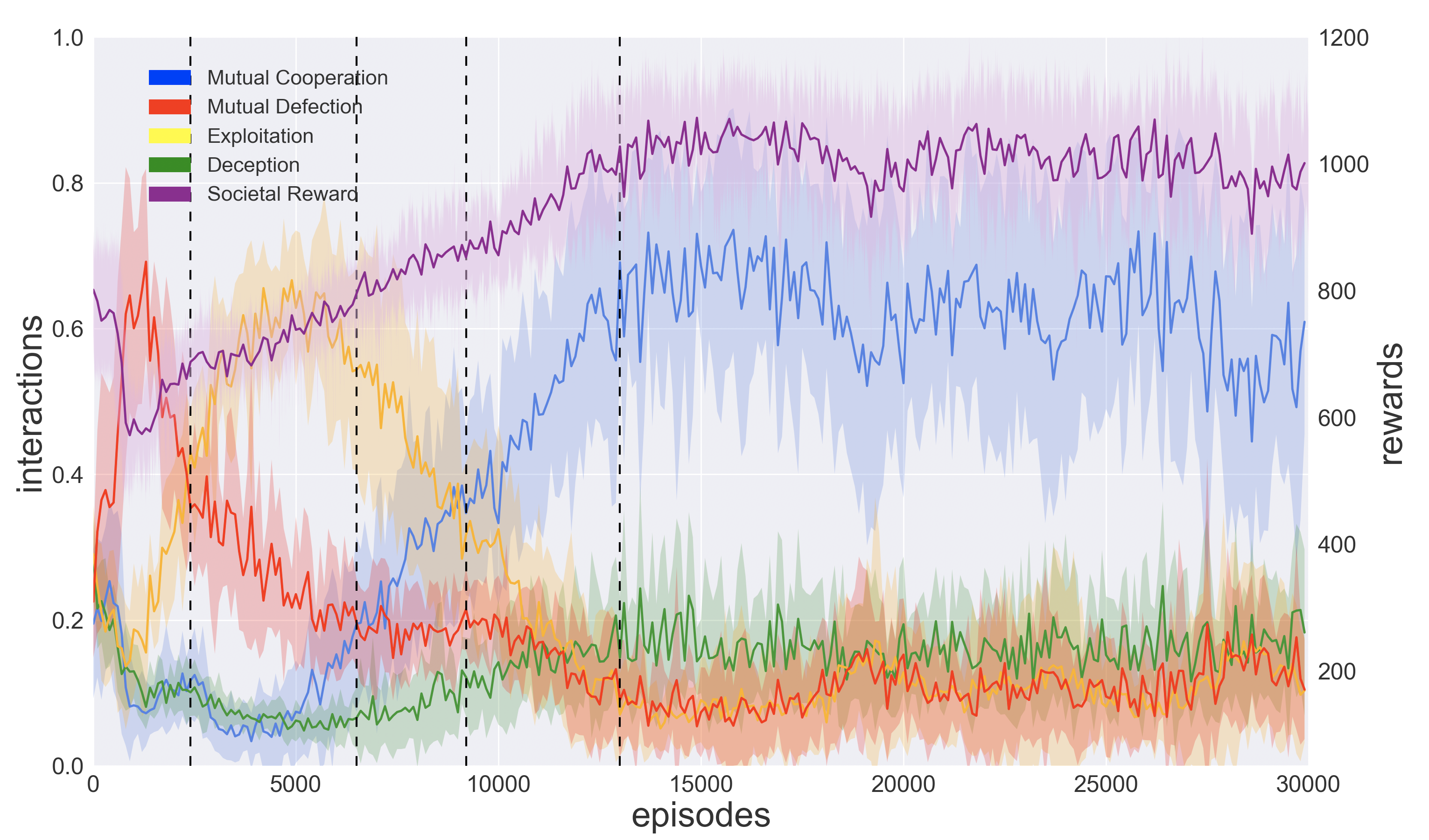}}
  \hspace{1cm}
  \subfigure[Random matching]{\includegraphics[width=0.95\columnwidth]{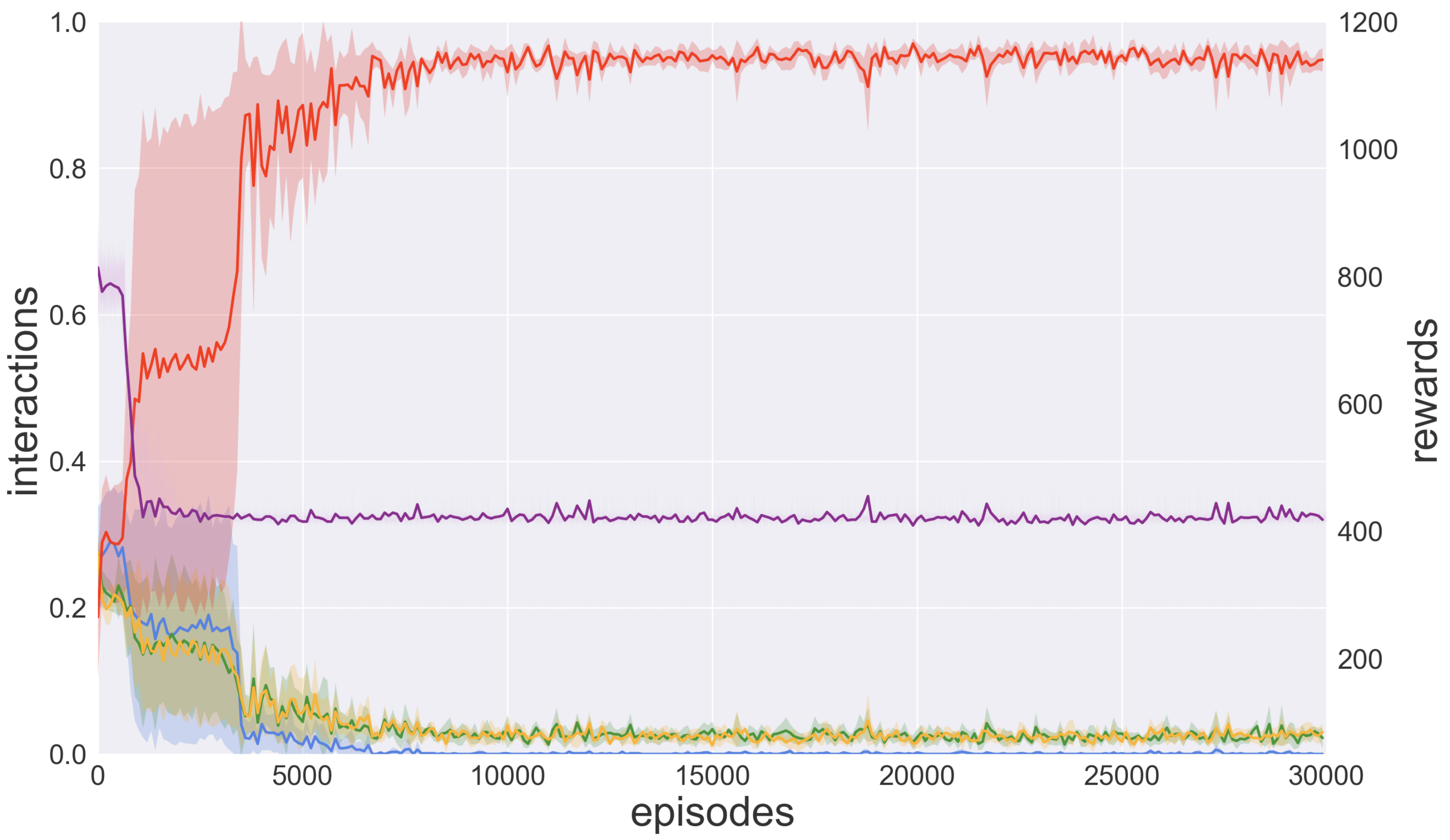}}

  \caption{The mean and standard deviation of the number of times an outcome occurs between two agents as a percentage over 14 runs of the simulation.  In our simulation, the dynamic of partner selection is essential for cooperation to emerge as a viable action which leads to a significantly higher global reward. With partner selection we observe distinct phases where certain strategies dominate. The start of each phase is marked with a dashed line. Eventually agents learn to predominantly cooperate after approximately 15,000 episodes. When pairing is random, agents quickly learn to defect and do not learn any cooperative strategy throughout the simulation.}
  \label{fig:emergence_of_coop}
\end{figure*}

In order to assess the emergent behavior associated with the dynamics of partner selection, it is important to be able to vary characteristics like the number of parameters and the agents' exploration rates for playing the Dilemma and selection. To give us this flexibility, each agent consists of two models: the selection model learns a policy $\pi_s$ to take action $a_s$ given $s_s$ while the dilemma model learns a policy $\pi_d$ to take action $a_d$ given $s_d$. During each episode, a trajectory $\tau = \{ s_s, a_s, r_s, s_d, a_d, r_d, ...\}$ was sampled from $\pi_s$ and $\pi_d$ and then each model was trained on the states and actions sampled from their respective policies. 

We parameterize each model using a neural network with one hidden layer. We test various network sizes between 32 and 256. We display the results for the case of a network size equal to 256 which were the most consistent and informative for understanding the development of strategy. Each agent employs an $\epsilon$-greedy policy with a fixed $\epsilon_d=0.05$ when playing the Dilemma and $\epsilon_s=0.1$ when selecting a partner. A discussion on $\epsilon$ is present in the Results section. Both models use the standard a discount rate $\gamma=0.99$.

\section{Results and Discussion}

\label{results}

In this section, we present results that indicate there is a significant impact on the development and emergence of cooperation as a result of partner selection. Alongside partner selection, we look at the roles of exploration and the types of learned strategies that promote a cooperative outcome. 

\subsection{Emergence of Cooperation through Partner Selection}

Training Q-agents in the two-player repeated Prisoner's Dilemma without partner selection results in both agents learning to defect every time. With random selection in a multi-agent population, agents converge to the same solution that is seen in the two-player case shown in Fig \ref{fig:emergence_of_coop}(b). When agents are given the ability to choose a partner within a population cooperation begins to emerge and the returns for the society steadily improve.

\begin{figure}[t]
    \begin{minipage}{1.0\columnwidth} 
        \centering

        \includegraphics[width=0.95\columnwidth]{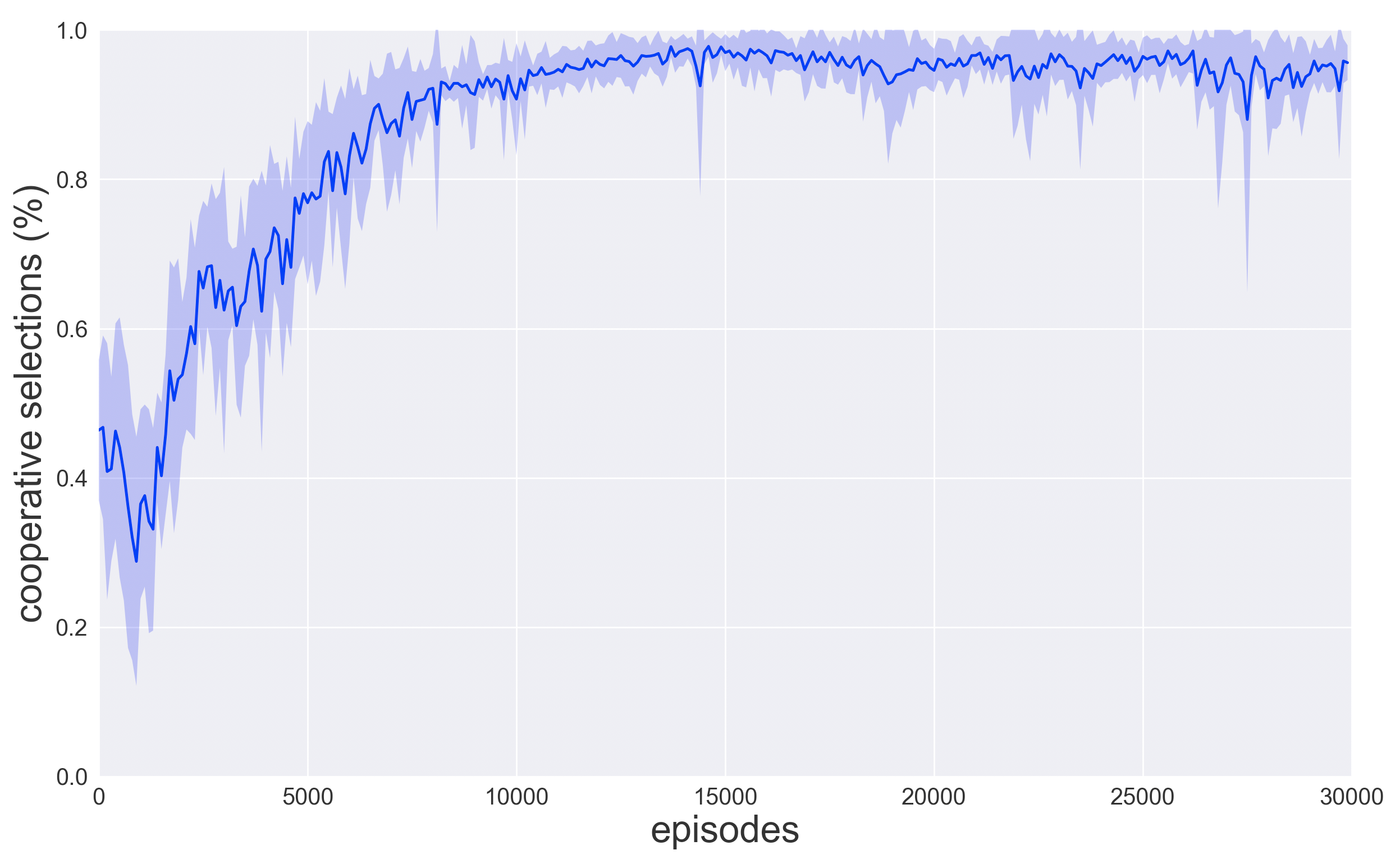}
        \caption{Percentage of agents selected who have cooperated in their last interaction.}
        \label{fig:partner_selection}
    \end{minipage}
\end{figure}

We observe four distinct phases characterized by different strategies of the agents. In particular, we identify four phases which we mark in Fig \ref{fig:emergence_of_coop}(a). We plot the number of agents that have selected a partner who has previously cooperated versus a partner who has previously defected in Fig \ref{fig:partner_selection}. In line with our expectations, at convergence, the partner of choice is a partner that has previously cooperated. Each of these four phases also coincides with a development in the strategy of the agents in the society which are as follows:

\begin{itemize}
    \item Phase 1: Agents learn to defect without partner selection (\textit{Rise in mutual defection}).
    \item Phase 2: Agents learn to pick a partner who has demonstrated cooperative behavior and defect against them (\textit{Rise in exploitation}). Agents pick a partner who has demonstrated cooperative behavior and cooperates with them (\textit{Rise in mutual cooperation}).
    \item Phase 3: Agents learn to retaliate (defect) against exploitative partners (\textit{Fall in exploitation)}. Agents learn to cooperate when selected by other cooperative agents (\textit{Rise in mutual cooperation}).
    \item Phase 4: We observe the emergence of a stable cooperative society with the majority of agents adopting a behavior that resembles equivalent retaliation, i.e., adoption of the Tit-for-Tat strategy.

\end{itemize}

\begin{figure*}[!ht]

  \subfigure[epsilon=0.1]{\includegraphics[width=0.95\columnwidth]{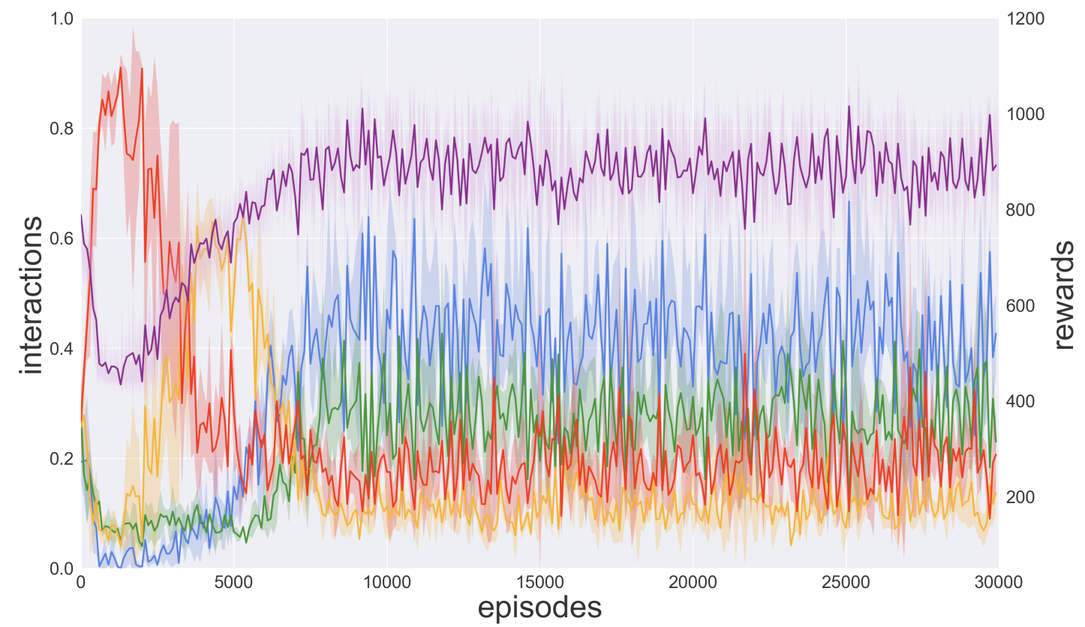}} \hspace{1.2cm}
  \subfigure[epsilon=0.01]{\includegraphics[width=0.95\columnwidth]{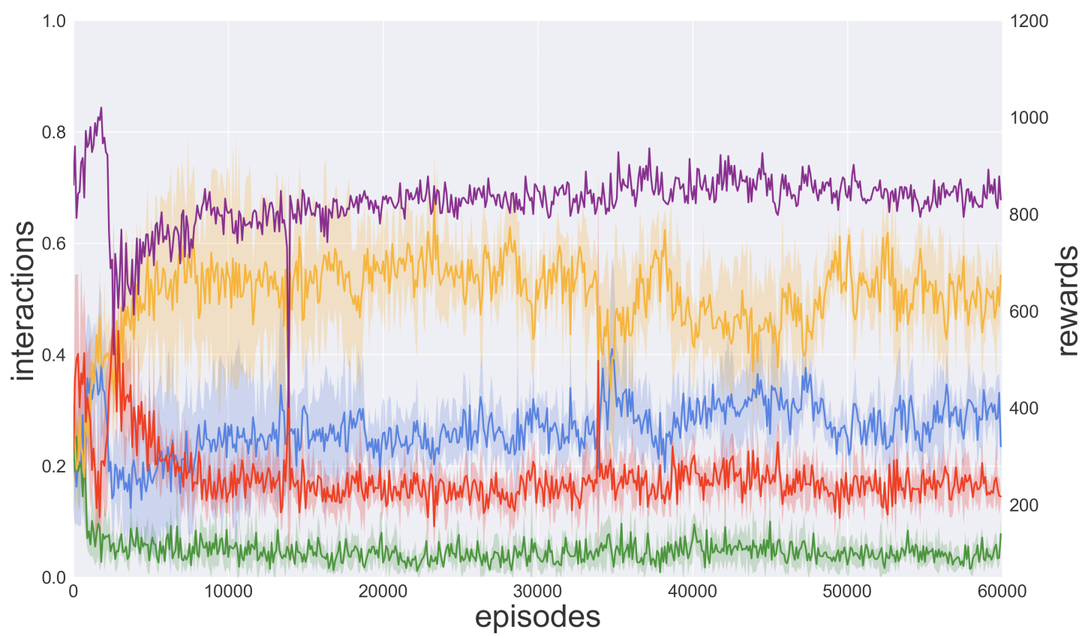}}

  \caption{Higher exploration results in the society going through the different phases quicker but agents demonstrate less overall cooperation. Lower exploration inhibits the transition to a cooperative phase. }
  \label{fig:exploration_rate}
\end{figure*}

\begin{figure*}[!ht]
    \includegraphics[width=2.1\columnwidth]{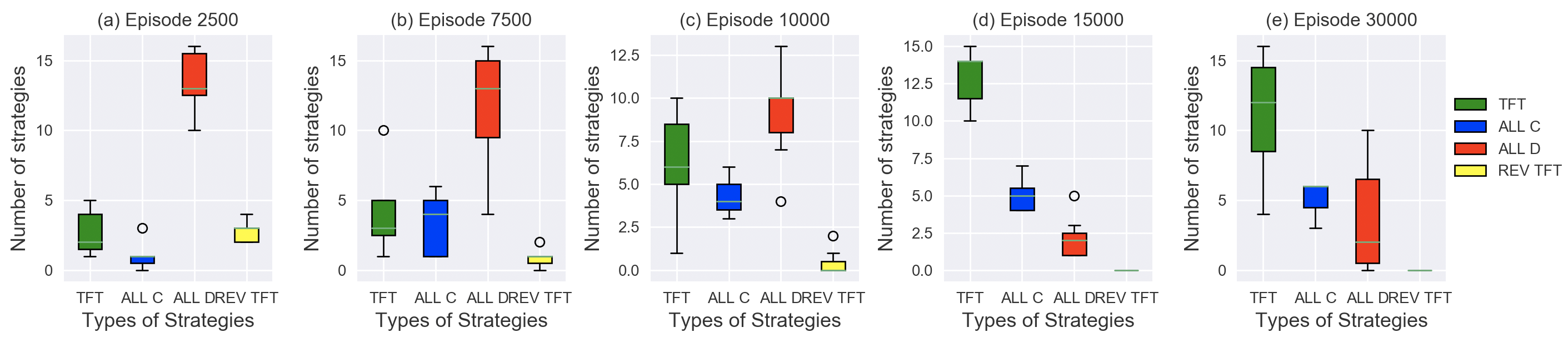}

    \caption{Box plots representing the number of agents that use each strategy during phase transitions. The rise in ALL C and TFT agents coincides with an improvement in partner selection. Once partner selection converges, the number of TFT agents rapidly increases which results in a more stable, cooperative society.}
    \label{fig:evol_strat}
\end{figure*}

When limiting the information an agent has of its opponent to just the action taken at the previous timestep we limit the space of strategies agents can learn to four strategies: (1) where agents always cooperate (ALL C), (2) where agents always defect (ALL D), (3) Tit-for-Tat (TFT) where agents copy the last action of their opponent, and (4) reverse Tit-for-Tat (revTFT), where agents play the opposite of their opponent's last action. To best understand the development of the agent strategies throughout the simulation, we capture both the outcomes of the agent encounters at every episode (Fig \ref{fig:emergence_of_coop}), the improvement in selection accuracy in (Fig \ref{fig:partner_selection}), and a box-plot of the strategies being used (\ref{fig:evol_strat}).

In the first phase, all agents quickly learn a defecting policy that causes a decrease in the society's cumulative reward. We can explain this outcome by observing that the agents have not yet learned how to select a partner efficiently. If the partner selection is uninformed, then each interaction can be treated independently and thus there are no future rewards associated with cooperating. Consequently, defecting is the preferred strategy for almost all agents. The partner selection accuracy of agents steadily improves, which facilitates the transition to the second phase.

The second phase begins at approximately 2,500 episodes into the simulation and is characterized by an increase in exploitation and a steep decrease in mutual defection. Agents have improved their capability to select cooperative partners, which increases the rewards associated with defecting. Although the outcome of the society is significantly different as a result of partner selection, the strategies used in the Dilemma are the same: agents primarily defect. What facilitates the transition to the next stage is the following: (1) agents who cooperate are selected to play more frequently than defecting agents (and, therefore, are given the opportunity to potentially receive rewards); and (2) with enough exploration, cooperation can be sufficiently rewarded and agents can start to learn to punish agents who would try to exploit them. 

This happens at approximately 5,000 episodes where there is a steep decline in exploitation. This leads to the third phase where there is a sufficient amount of cooperators such that they can be easily selected and exploiters begin to be punished. In episode 7,500 there is a steady increase in the number of encounters resulting in mutual cooperation. Tracking the agent strategies (Fig \ref{fig:evol_strat}) reveals that this can be attributed to the rise in TFT. With an increase in the number of TFT agents, defecting agents exploit others less successfully as an agent who has previously cooperated is more likely to be a TFT agent than an ALL C agent resulting in encounters leading to mutual defection. At the 10,000 episode mark, mutual cooperation overtakes exploitation as the most common outcome of two agents interacting that is observed in the society. Looking at Fig \ref{fig:partner_selection}, this also coincides with the convergence of partner selection as 90\% of agents who are selected to play the Dilemma have cooperated in their last interaction. After 15,000 training episodes, the society enters a phase of where a significant majority of interactions result in mutual cooperation and the society has mostly stabilized. This is also where the society achieves its highest cumulative reward. It is expected that at this point, where the number of TFT agents is at its highest, is where the number of ALL D agents is at its lowest. The number of ALL C agents stays relatively consistent from phase 2 throughout to phase 4, although it increases when the ALL D strategy is no longer popular. 

The impact of TFT agents here can be clearly seen as they significantly limit the effectiveness of the defecting strategy. As long as agents use ALL D, the best response strategy is to employ TFT. Furthermore, when more agents use TFT, the more rewards TFT agents will receive during an episode which further pushes agents to employ it as a strategy. We can see that once agents are able to select cooperative partners, the TFT strategy acts as a norm by inducing other agents to play the same strategy which produces the most stable iteration of the society.

\subsection{Analyzing Examples From Simulation}

\begin{figure*}[!h]
	\centering
    \includegraphics[width=2.0\columnwidth]{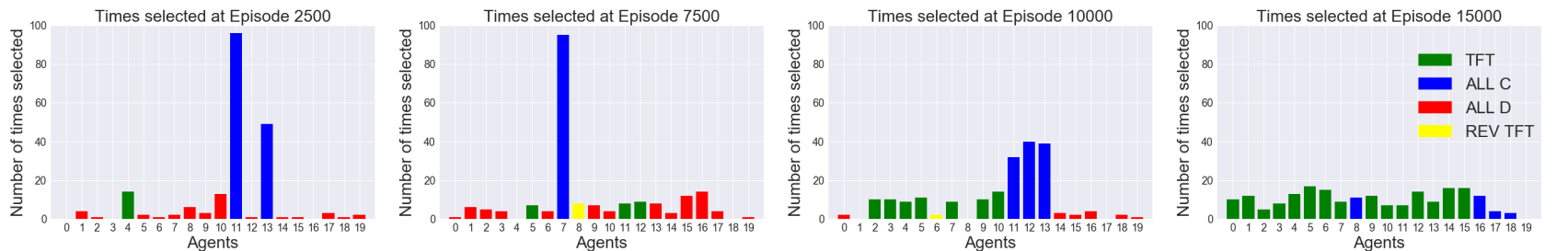}
   
    \caption{With every agent searching for a partner that indicates that they will cooperate, the agent that is most commonly selected are ones that use ALL C. Early on in the training, ALL D is quickly adopted by most agents while reverse TFT is phased out. As agents learn TFT, the distribution of selections flattens. As more agents play TFT, less agents play ALL D.}
    \label{fig:selection_dist}
\end{figure*}

\begin{figure*}[!h]
	\centering
    \includegraphics[width=2.0\columnwidth]{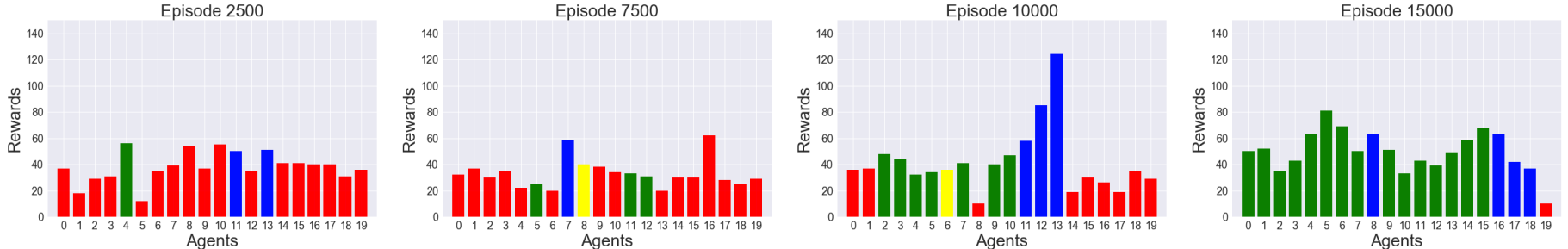}
    \caption{In the early stages of the simulation and very few cooperative agents, the spread of rewards per episode is fairly even. Despite being selected the vast majority of the time, ALL C agents receive similar rewards per episode. As other agents learn TFT, the amount of reward that ALL C agents receive significantly increases, while it decreases for ALL D agents. Eventually, when the majority of agents play TFT, the average rewards per agent increases.}
    \label{fig:rewards_per_episode}
\end{figure*}

\begin{figure*}[!h]
	\centering
    \includegraphics[width=2.0\columnwidth]{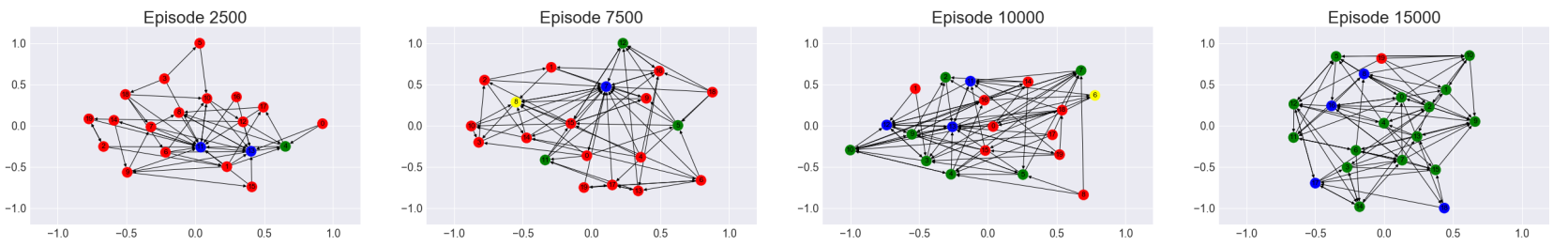}
    \caption{Degree centrality plots of 20 agents in a society during training. Each node represents an agent, while each directed edge represents an interaction between two agents. Agents with more interactions between each other have shorter edges. Axes simply represent a normalized node position centered on the origin.}
    \label{fig:network}
\end{figure*}

Tracking the returns of each agent alone is insufficient to capture the state of the society and the group's behavior. For a more granular understanding, we further plot examples from the phases from a simulation to show the number of times each agent is selected and the strategies they use (Fig \ref{fig:selection_dist}), the individual reward they receive (Fig \ref{fig:rewards_per_episode}), and the interactions represented as a network (Fig \ref{fig:network}). We specifically consider examples of interactions between the phase transitions described in the previous section.

In episode 2,500, rewards are evenly spread across agents, however, selections are heavily skewed in favor of cooperative agents. Nevertheless, they receive very similar rewards to other agents using TFT or ALL D strategies. ALL D agents that are picked more often do not receive significantly more rewards as they are most often being selected by agents who are also defecting meaning little reward is gained from those encounters. In episode 7,500, we observe more agents using a TFT strategy. However, the two most successful strategies in this phase are ALL D (agent 16) and ALL C (agent 7) both of whom benefit from more effective partner selection.

In phase 3, surprisingly, we observe the largest amount of inequality. Due to the growing number of TFT agents, agents using ALL C strategy receive significantly more reward while agents that use ALL D and receive the smaller rewards on average. The amount of reward the defecting agents receive is lower due to a combination of not being selected by other agents, the number of TFT agents in the population (making it difficult for them to find cooperative agents to exploit), as well as cooperative agents being able to reliably cooperate with one another. With only a single available history, the ALL D agents cannot distinguish between an ALL C agent and a TFT agent that has recently cooperated. Interestingly, when a TFT agent punishes an ALL D agent (or faces a TFT or ALL C agent who has recently explored defection), the TFT agent suffers a cost in the next round as their chance of getting selected is reduced despite having behaved in a cooperative way by punishing a defector. This explains why ALL C receives the highest overall reward. As ALL C always cooperates, it does not face this penalty but still benefits from the regulatory behavior provided by other TFT agents. Nevertheless, we can see that the strategy that is adopted by most agents is surprisingly not the strategy that necessarily generates the highest reward in each episode.
 
\subsection*{The Role Of Exploration}

In RL, it is necessary to give agents the opportunity to explore the environment to find an optimal strategy. In multi-agent systems when the agents are continuously developing new strategies, there is a need to explore constantly in order to react to the shift in environment dynamics. However, agents exploring can impact the learning of other agents. Too much exploration at the wrong time can inhibit the convergence to certain behaviors. As an example, if a single agent has a cooperative strategy when all the other agents have a defecting strategy, then exploration benefits the cooperative strategy. On the other hand, when all agents behave cooperatively, exploration instead becomes costly for cooperators. In our work we have suggested that, in order to achieve a cooperative outcome, the agent strategy undergoes four stages. We can see that increasing the base exploration rate affects the rate at which these strategies are learned, however, it also significantly affects the overall society's return. In Fig \ref{fig:exploration_rate}(a) a higher exploration rate results in an overall less cooperative outcome and, therefore, less cumulative reward. Similarly, in Fig \ref{fig:exploration_rate}(b), if the exploration rate is too low, it makes it difficult for agents to adjust their strategies and, despite doubling the number of episodes, the agents are not able to reach a cooperative solution. 

Although more exploration facilitates a quicker transition between phases, it also jeopardizes the stability and gains of the final outcome. We should therefore be mindful in how we determine the exploration strategy that our agents employ in the presence of other learners.

\subsection*{Inclusion and Exclusion to Promote Norm-Inducing Behavior}

The impact of partner selection happens in two steps that can be viewed as inclusive acts and exclusive acts. Firstly, agents need to be able to distinguish between different types of agents. Cooperators need to be included to allow cooperation to compete with defecting strategies for reward. Secondly, agents need to be able to learn or adapt their playing strategy given a selection strategy that leads to the exclusion of and retaliation against defectors. We view this behavior as \textit{norm-inducing} as it forces agents to conform to an existing strategy that has been adopted by the majority of the agents. In order for the norm to persist, the TFT agents need to interact with a mix of both ALL C and ALL D agents and, while it regulates the behavior of the other agents, the emergence and presence of TFT behavior itself is not sustained by it. Although agents that use an ALL C strategy tend to receive the highest rewards, without other agents employing a TFT strategy, the stability of a cooperative society is threatened. Our experiments suggest that the emergence of cooperation is contingent on having a sufficient number of agents learning to exclude bad behavior even if they share some of the associated cost.
 
In Fig \ref{fig:network}, we display centrality plots from network analysis to visualize partner selection with nodes positioned according to Fruchterman-Reingold force-directed algorithm \cite{fruchterman1991graph}. In these plots, two nodes are close if there are more interactions between them. As partner selection becomes more effective in terms of reward with more training episodes, the network gets tighter with ALL C nodes having a higher degree centrality and influence. In the examples above, ALL D agents are prominent outside the core of the network while agents who cooperate have higher degree centrality. As training progresses, the connections are distributed more evenly as more agents adopt equivalent-retaliation (TFT) strategy and become viable partners for cooperation. With more TFT agents, it is harder for ALL D agents to find ALL C agents to exploit which decreases the amount of connections between those two node types. This causes a change in the network structure where there is less likely to be just one or two highly influential nodes. Finally, as mutual cooperation overtakes exploitation and the society stabilizes, the network connections become more evenly spread with nodes having similar degree centrality.

\section{Conclusions}

We have presented an analysis of the effect of partner selection on the emergence of cooperation. We have shown that agents can learn to cooperate in the repeated Prisoner's Dilemma using multi-agent reinforcement learning. We have outlined four key phases that begin with defection but, as partner selection improves, the agents achieve a cooperative outcome. Although subject to certain environment variables, such as reward and the amount of information available to each agent, partner selection allows for a cooperative strategy to be sustained early on in the simulation when the majority of agents defect. This facilitates a development of a strategy that learns to interact efficiently with both cooperative and defective agents akin to Tit-for-Tat. The stability of a cooperative outcome is dependent on how successfully agents are able to use information about another agent's history to inform partner selection and their strategy in the Dilemma. Although TFT agents tend to receive less reward than fully cooperative agents at certain stages, the presence of this strategy is required for cooperation to emerge at all as it regulates the exploitative behavior of defecting agents. The results of this work demonstrate the importance of understanding how agents can learn to interact with each other in an environment beyond internal agent or reward function design that can help produce optimal cooperative outcomes and provides insight into the emergence of cooperation in social dilemmas.

\bibliography{references}
\bibliographystyle{aaai}

\end{document}